\documentclass[preprint,12pt]{elsarticle}

\usepackage{amssymb}
\usepackage{amsmath}

\journal{Nuclear Physics B}

\begin{document}

\begin{frontmatter}



\title{KEMPIC-3D: A transparent and extensible electromagnetic Particle-in-Cell framework for kinetic plasma simulations}

\author[label1]{Jes\'us E. L\'opez}
\author[label1,label2,label3]{Keren C. Vanegas}
\author[label1]{Eduardo A. Orozco-Ospino}

\affiliation[label1]{
    organization={Universidad Industrial de Santander},
    city={Bucaramanga},
    postcode={680002},
    state={Santander},
    country={Colombia}
}

\affiliation[label2]{
    organization={ELI Beamlines Facility, The Extreme Light Infrastructure ERIC},
    addressline={Za Radnicí 835},
    city={Doln\'i B\v{r}e\v{z}any},
    postcode={25241},
    country={Czech Republic}
}

\affiliation[label3]{
    organization={Czech Technical University in Prague, Faculty of Nuclear Sciences and Physical Engineering,},
    addressline={B\v{r}ehov\'a 7},
    city={Prague},
    postcode={11519},
    country={Czech Republic}
}

\begin{abstract}
KEMPIC-3D, a fully electromagnetic three-dimensional particle-in-cell (PIC) code, has been developed in C/C++ and systematically verified for the self-consistent simulation of kinetic plasma phenomena. The numerical framework combines a staggered Yee finite-difference time-domain solver for Maxwell's equations with a relativistic Boris particle pusher, trilinear field interpolation, and charge-conserving current deposition. Its modular architecture emphasizes algorithmic transparency and extensibility, allowing the principal numerical routines to be directly inspected and modified to incorporate additional physical models and computational capabilities.

The implementation is assessed through a hierarchy of benchmark problems addressing both the individual numerical components and the self-consistent particle--field coupling. The particle pusher is verified against analytical solutions for relativistic charged-particle motion, while the electromagnetic field solver is evaluated through guided-wave propagation and grid-convergence analysis. The complete PIC algorithm is subsequently verified through plasma oscillations, global energy conservation, and discrete charge continuity, demonstrating the numerical accuracy, conservation properties, and self-consistency of the coupled formulation.

The capabilities of KEMPIC-3D are further demonstrated through a representative simulation of laser-driven plasma wakefield generation in an underdense plasma. The simulations reproduce the characteristic nonlinear wakefield structure and exhibit stable numerical behavior across different particle discretizations without additional spatial filtering. Together, these results establish KEMPIC-3D as a reliable, transparent, and extensible computational framework for multidimensional kinetic plasma simulations. The source code will be made publicly available to facilitate reproducibility, independent use, and further development by the plasma physics community.
\end{abstract}




\end{frontmatter}



\section{Introduction}\label{introduction}
Kinetic plasma phenomena involve nonlinear interactions between charged particles and electromagnetic fields over a wide range of spatial and temporal scales. In many regimes, fluid descriptions are insufficient to represent wave--particle interactions, non-equilibrium distribution functions, and microscopic collective processes. Particle-in-cell (PIC) methods provide a self-consistent kinetic description by coupling the numerical solution of Maxwell's equations to the motion of computational particles and have become one of the principal approaches in computational plasma physics~\cite{Dawson1983,BirdsallLangdon2004,HockneyEastwood1988}. Their applications include plasma-based acceleration, laser--plasma interactions, magnetic-confinement plasmas, space physics, and astrophysical environments~\cite{Fonseca2002,Arber2015,Derouillat2018,Vay2018,Lehe2016}.

Since the early development of plasma particle simulations, PIC methods have evolved into mature computational tools and have motivated the development of several sophisticated simulation frameworks. Representative examples include OSIRIS~\cite{Fonseca2002}, EPOCH~\cite{Arber2015}, Smilei~\cite{Derouillat2018}, WarpX~\cite{Vay2018}, and FBPIC~\cite{Lehe2016}. These frameworks provide advanced numerical models and high-performance implementations for large-scale kinetic plasma simulations. Nevertheless, the development of independent PIC implementations remains valuable for studying the numerical method in detail, retaining direct control over the computational framework, and incorporating algorithms or physical models tailored to specific applications.

In this context, KEMPIC-3D has been developed with a different emphasis from large-scale high-performance PIC platforms. Rather than targeting high-performance computing as its primary design objective, the code prioritizes algorithmic transparency, modularity, and extensibility, while incorporating a lightweight parallelization strategy for multidimensional simulations. The principal numerical routines are organized so that their implementation can be directly inspected and readily modified, facilitating the incorporation of additional physical models, numerical methods, and diagnostic capabilities. This design philosophy provides a versatile and accessible computational environment for kinetic plasma research.

The reliability of a PIC implementation depends critically on the consistency of the coupling between particles and fields. Numerical errors introduced during field interpolation, particle pushing, or charge and current deposition may produce violations of the discrete continuity equation, inconsistency with Gauss's law, or artificial energy growth~\cite{BirdsallLangdon2004,HockneyEastwood1988,VillasenorBuneman1992,Esirkepov2001}. Systematic verification is therefore required before the code is applied to complex plasma configurations. This process should assess the accuracy and convergence of the individual numerical components as well as the conservation properties and characteristic physical responses of the fully coupled particle--field system.

Earlier versions of KEMPIC-3D have already been employed in published studies of microwave-driven plasma wakefields. In a first application, three-dimensional electromagnetic PIC simulations were used to investigate the formation and structure of plasma wakefields driven by short microwave pulses in rectangular plasma-filled waveguides~\cite{lopez2025particle}. A subsequent study used the same computational framework to investigate the acceleration of externally injected electrons by these wakefields, including the effects of injection phase, initial velocity, transverse dynamics, and space charge~\cite{lopez2026numerical}. While these studies focused on specific plasma-acceleration problems, the present work focuses on KEMPIC-3D itself, providing a systematic description and verification of its numerical implementation, together with its public release to facilitate broader use and further development.

In this work, KEMPIC-3D, a fully electromagnetic, three-dimensional PIC code implemented in C++, is presented and systematically verified. The numerical framework combines the finite-difference time-domain Yee scheme~\cite{Yee1966} for solving Maxwell's equations with the relativistic Boris particle pusher~\cite{Boris1970}, charge-conserving current deposition based on the ZigZag method~\cite{Umeda2003}, and trilinear particle--field interpolation.

The numerical implementation is assessed progressively through verification tests of the particle pusher, the electromagnetic field solver, and the self-consistent particle--field coupling. The Boris integrator is examined under uniform electric and magnetic fields, whereas the Yee solver is evaluated through electromagnetic propagation in a rectangular waveguide. The self-consistent particle--field coupling is subsequently assessed through plasma oscillations, plasma-frequency recovery, energy conservation, and the discrete continuity equation. Finally, the excitation of a nonlinear laser-driven plasma wakefield is presented as a representative application of the computational framework. Together, these results establish the numerical consistency of KEMPIC-3D and demonstrate its capabilities for multidimensional kinetic plasma simulations.
\section{Numerical model}\label{numerical_model}
The particle-in-cell (PIC) method is a hybrid particle--mesh approach in which the plasma is represented by computational particles evolving in a Lagrangian framework, while the electromagnetic fields are defined on a fixed Eulerian mesh~\cite{BirdsallLangdon2004,HockneyEastwood1988}. The interaction between these two descriptions constitutes the essence of the PIC methodology: particles act as the source of the electromagnetic fields through their charge and current densities, whereas the updated fields determine the subsequent particle dynamics. This mutual coupling gives rise to a fully self-consistent description of plasma evolution, enabling the simulation of collective kinetic phenomena over a broad range of spatial and temporal scales.

The numerical framework developed in this work follows this self-consistent particle--field paradigm and combines a set of well-established numerical algorithms to accurately solve the coupled evolution of particles and electromagnetic fields. Particular emphasis has been placed on ensuring numerical consistency, stability, and the preservation of the fundamental physical properties required for reliable, fully electromagnetic kinetic plasma simulations.
\subsection{Governing equations and PIC formulation}
The evolution of collisionless plasmas is governed by the coupled Vlasov--Maxwell system, which provides a self-consistent kinetic description of charged particles and electromagnetic fields~\cite{BirdsallLangdon2004,HockneyEastwood1988}. Rather than describing individual particle trajectories, the kinetic approach represents each plasma species $\alpha$ through a distribution function $f_\alpha(\mathbf{r},\mathbf{p},t)$ defined in the six-dimensional phase space. In the absence of collisions, its evolution is governed by the relativistic Vlasov equation,

\begin{equation}
    \frac{\partial f_{\alpha}}{\partial t}
    + \frac{\mathbf{p}}{\gamma m_{\alpha}}\cdot\nabla_{\mathbf r}f_{\alpha}
    + q_{\alpha}
    \left(
        \mathbf E+
        \frac{\mathbf p}{\gamma m_{\alpha}}\times\mathbf B
    \right)
    \cdot
    \nabla_{\mathbf p}f_{\alpha}
    =0,
    \label{EQ_Vlasov}
\end{equation}

where $\gamma=(1-v^2/c^2)^{-1/2}$ is the Lorentz factor, while $\mathbf{E}$ and $\mathbf{B}$ denote the electric and magnetic fields acting on the plasma. The electromagnetic fields evolve simultaneously according to Maxwell's equations,

\begin{align}
\nabla\cdot\mathbf{E} &= \frac{\rho}{\varepsilon_0},\label{eq_gauss_E_law}\\
\nabla\cdot\mathbf{B} &= 0,\label{eq_gauss_B_law}\\
\nabla\times\mathbf{E} &= -\frac{\partial\mathbf{B}}{\partial t},\label{eq_faraday_law}\\
\nabla\times\mathbf{B} &= \mu_0\mathbf{J}
+\mu_0\varepsilon_0
\frac{\partial\mathbf{E}}{\partial t},
\label{eq_ampere_maxwell_law}
\end{align}

where the charge density $\rho$ and current density $\mathbf{J}$ provide the coupling between particles and electromagnetic fields,

\begin{align}
\rho(\mathbf r,t)
&=
\sum_\alpha
q_\alpha
\int
f_\alpha(\mathbf r,\mathbf p,t)
\,d^3\mathbf p,
\\
\mathbf J(\mathbf r,t)
&=
\sum_\alpha
q_\alpha
\int
\frac{\mathbf p}{\gamma m_\alpha}
f_\alpha(\mathbf r,\mathbf p,t)
\,d^3\mathbf p.
\end{align}

\begin{figure}[t]
    \centering
    \includegraphics[width=\linewidth]{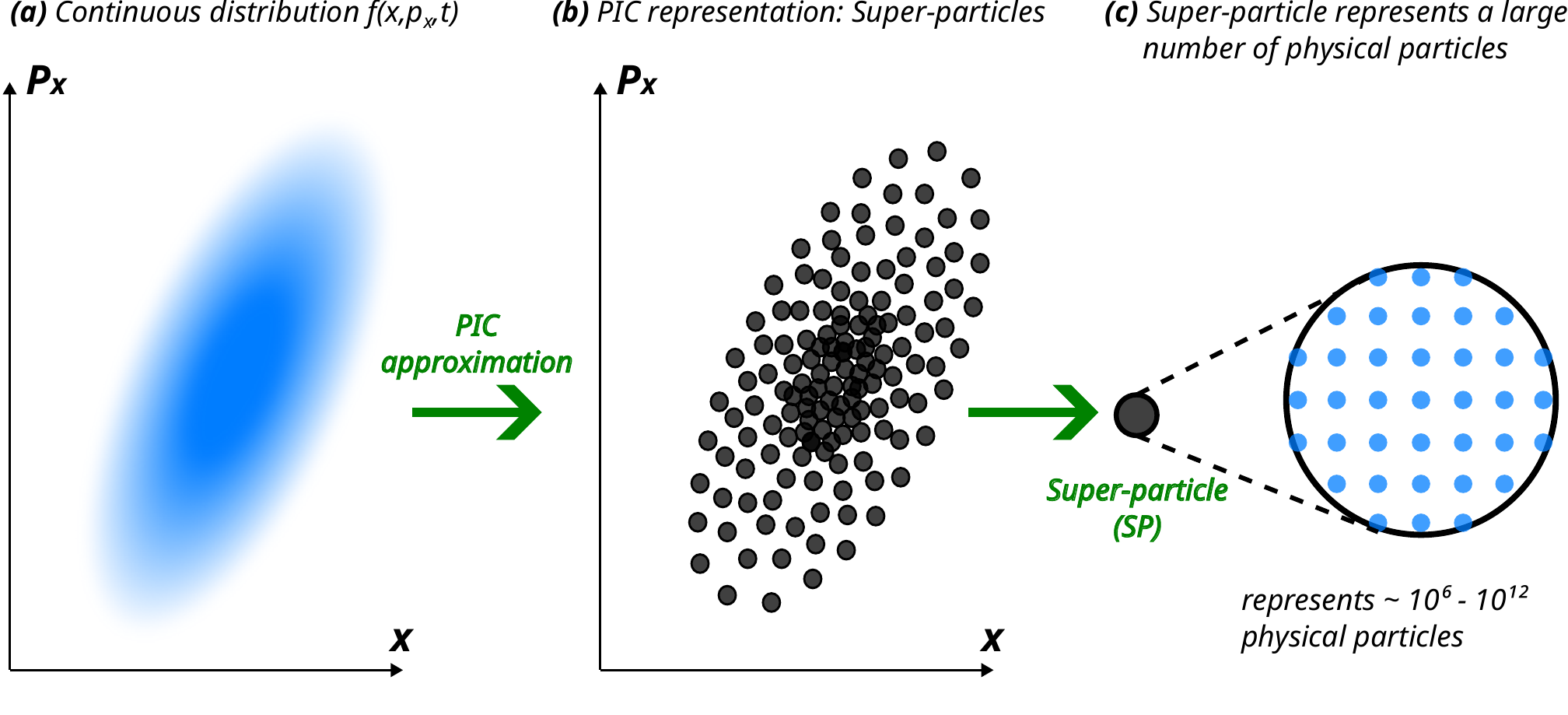}
    \caption{Discretization of the distribution function in phase space using the particle-in-cell (PIC) method. (a) Continuous phase-space distribution function $f(x,p_x,t)$. (b) PIC representation, where the continuous distribution is approximated by a finite set of super-particles (SPs) sampling the distribution function in phase space. (c) Each super-particle represents an ensemble of physical particles occupying a small region of phase space and evolving collectively as a single computational element.}
    \label{FIG_PIC_SP}
\end{figure}

Although Eq.~(\ref{EQ_Vlasov}) provides the continuous kinetic description of the plasma, directly solving the six-dimensional distribution function is computationally prohibitive for most practical applications. The particle-in-cell (PIC) method overcomes this difficulty by discretizing the continuous distribution function into a finite set of computational elements known as super-particles (SPs)~\cite{BirdsallLangdon2004,HockneyEastwood1988}. Each SP represents an ensemble of physical particles occupying a small region of phase space and collectively samples the evolution of the distribution function. In practice, every SP carries a finite weight corresponding to the number of physical particles it represents, allowing the continuous distribution to be approximated using a computationally tractable number of degrees of freedom. Accordingly,

\begin{equation}
f_\alpha(\mathbf r,\mathbf p,t)
=
\sum_{sp=1}^{N_{SP}}
f_{sp}(\mathbf r,\mathbf p,t),
\label{EQ_fPIC}
\end{equation}

where $N_{SP}$ is the total number of super-particles representing species $\alpha$. Figure~\ref{FIG_PIC_SP} illustrates this approximation by comparing the continuous distribution function with its discrete representation in terms of SPs.

Within this formulation, the kinetic evolution of the plasma is replaced by the evolution of the SPs. Rather than solving Eq.~(\ref{EQ_Vlasov}) directly, the PIC method advances the position and momentum of every SP according to the relativistic Newton--Lorentz equations,

\begin{align}
\frac{dN_p}{dt} &=0,
\\
\frac{d\mathbf r_{sp}}{dt}
&=
\mathbf v_{sp},
\\
\frac{d(\gamma m_{sp}\mathbf v_{sp})}{dt}
&=
q_{sp}
\left(
\mathbf E_{sp}
+
\mathbf v_{sp}\times\mathbf B_{sp}
\right),
\label{EQ_motion_SP}
\end{align}

where $N_p$ denotes the number of physical particles represented by each SP, which remains constant throughout the simulation, while $m_{sp}=N_pm_\alpha$ and $q_{sp}=N_pq_\alpha$ are the corresponding mass and charge of the SP.

The electric and magnetic fields acting on each SP $\mathbf{E}_{sp}$ and $\mathbf{B}_{sp}$, are obtained by interpolating the grid quantities to the particle position through the adopted particle shape function~\cite{BirdsallLangdon2004,HockneyEastwood1988}. Conversely, the charge and current carried by the SPs are deposited onto the computational mesh, providing the source terms required to advance Maxwell's equations. Consequently, the particle and field solvers become mutually coupled through a continuous exchange of information, giving rise to the self-consistent feedback mechanism that characterizes fully electromagnetic PIC simulations. This continuous particle--field interaction constitutes the fundamental self-consistent cycle of the PIC method and forms the computational backbone of the numerical framework developed in this work.
\subsection{Electromagnetic PIC algorithm}
The numerical solution of the coupled Vlasov--Maxwell system is obtained through an iterative particle--field cycle in which the electromagnetic fields and the super-particles are advanced self-consistently in time. The implementation follows the conventional leapfrog time staggering employed in fully electromagnetic PIC methods~\cite{BirdsallLangdon2004,HockneyEastwood1988}, in which the electric field, magnetic field, particle positions, and particle momenta are defined at staggered time levels to achieve second-order temporal accuracy and improved numerical stability.

At each time step, the magnetic field is first advanced by half a time step using Faraday's law, after which the electric field is interpolated to the particle positions together with the time-centered magnetic field. These fields are then employed to evaluate the Lorentz force and advance the particle momenta and positions through the relativistic equations of motion. The updated particle trajectories are subsequently used to deposit the charge and current densities onto the computational mesh, providing the source terms required to complete the electromagnetic field update through Ampère--Maxwell's law. Finally, additional numerical operations, including the application of boundary conditions, compensated smoothing procedures, and, when required, the moving-window technique, are performed before the next iteration begins.

\begin{figure}[h]
    \centering
    \includegraphics[width=\linewidth]{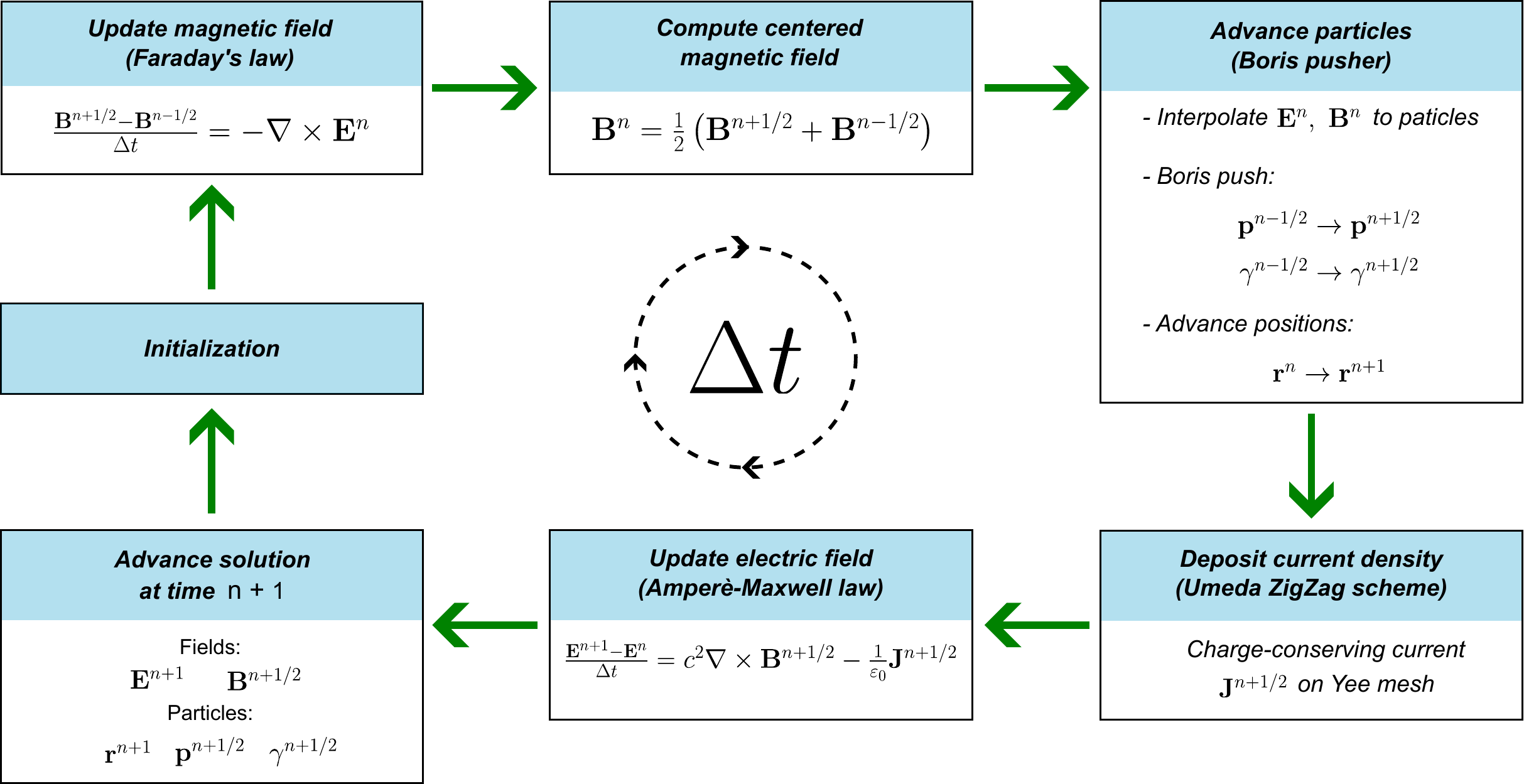}
    \caption{Computational workflow implemented in KEMPIC-3D during one simulation time step. The algorithm advances the solution through the following sequence: (1) update of the magnetic field using Faraday's law, (2) temporal centering of the magnetic field required by the Boris particle pusher, (3) advancement of particle positions and momenta using the relativistic Boris algorithm, (4) charge-conserving current deposition onto the Yee mesh using the ZigZag scheme of Umeda \textit{et al.}, (5) update of the electric field through the discretized Ampère--Maxwell equation, and (6) updated electromagnetic fields and particle quantities that provide the input for the next PIC iteration.}
    \label{FIG_PIC_algorithm}
\end{figure}

This sequence of operations establishes the self-consistent feedback mechanism that characterizes the electromagnetic PIC method: particle motion generates the charge and current densities that drive Maxwell's equations, while the updated electromagnetic fields determine the subsequent particle dynamics. Repeating this cycle until the prescribed final simulation time yields the complete self-consistent evolution of the particle distribution and electromagnetic fields. A schematic overview of the computational workflow implemented in the present code is shown in Fig.~\ref{FIG_PIC_algorithm}, while the numerical techniques employed in each stage are described in the following subsections.

\subsection{Electromagnetic field solver}
The self-consistent evolution of the electromagnetic fields is obtained by solving Maxwell's equations on a structured Cartesian mesh using the finite-difference time-domain (FDTD) scheme proposed by Yee~\cite{Yee1966}. Owing to its second-order accuracy, explicit formulation, and excellent compatibility with electromagnetic particle-in-cell simulations, the Yee scheme has become one of the most widely used electromagnetic field solvers in modern PIC codes~\cite{BirdsallLangdon2004,Taflove2005,Derouillat2018}. In the present work, this formulation was selected because of its numerical robustness, computational efficiency, and ability to preserve the discrete structure of Maxwell's equations on regular grids.

The Yee scheme advances only the curl equations, namely Faraday's law, Eq.~(\ref{eq_faraday_law}), and the Ampère--Maxwell equation, Eq.~(\ref{eq_ampere_maxwell_law}), while Gauss's law, Eq.~(\ref{eq_gauss_E_law}), and the divergence-free condition for the magnetic field, Eq.~(\ref{eq_gauss_B_law}), are not solved explicitly. Instead, these constraints are preserved throughout the simulation provided that they are satisfied initially and that the deposited charge and current densities satisfy the discrete continuity equation~\cite{Verboncoeur2005,Esirkepov2001},

\begin{equation}
\frac{\partial \rho}{\partial t}
+
\nabla\cdot\mathbf J
=
0,
\label{eq_continuity_equation}
\end{equation}

The condition $\nabla\cdot\mathbf B=0$ is naturally maintained because the discrete divergence of the discrete curl vanishes identically within the Yee discretization~\cite{Yee1966,Taflove2005}. Consequently, preserving the discrete form of Gauss's law relies not only on the field update equations but also on the use of a charge-conserving current deposition scheme within the PIC cycle.

The FDTD formulation combines a staggered spatial discretization with a leapfrog time integration scheme. The electric and magnetic field components are stored at different locations within each computational cell and are staggered in time by half a time step. Specifically, the electric field is evaluated at integer time levels $t^n$, whereas the magnetic field is evaluated at half-integer time levels $t^{n+\frac12}$. This arrangement enables centered second-order finite differences in both space and time while providing a natural discretization of the curl operators appearing in Maxwell's equations.

The corresponding spatial staggering places each component of the electric and magnetic fields on its own staggered sub-grid. This arrangement allows each spatial derivative to be evaluated using neighboring field components, resulting in a compact local update stencil. As a result, the field solver is explicit: the electromagnetic fields at the next time level depend only on neighboring field values available from the previous update together with the current density deposited by the particles.

For example, the update equation for the $x$-component of the magnetic field is given by

\begin{equation}
\begin{aligned}
B^{\,n+\frac12}_x(i,j+\tfrac12,k+\tfrac12)
&=
B^{\,n-\frac12}_x(i,j+\tfrac12,k+\tfrac12)
\\
&-
\frac{\Delta t}{\Delta y}
\left[
E^{\,n}_z(i,j+1,k+\tfrac12)
-
E^{\,n}_z(i,j,k+\tfrac12)
\right]
\\
&+
\frac{\Delta t}{\Delta z}
\left[
E^{\,n}_y(i,j+\tfrac12,k+1)
-
E^{\,n}_y(i,j+\tfrac12,k)
\right],
\end{aligned}
\label{EQ_Yee_Bx}
\end{equation}

whereas the electric field is updated according to the discretized Ampère--Maxwell equation. As an example, the $x$-component evolves as

\begin{equation}
\begin{aligned}
E^{\,n+1}_x(i+\tfrac12,j,k)
&=
E^{\,n}_x(i+\tfrac12,j,k)
\\
&+
\frac{\Delta t}{\Delta y}
\left[
B^{\,n+\frac12}_z(i+\tfrac12,j+\tfrac12,k)
-
B^{\,n+\frac12}_z(i+\tfrac12,j-\tfrac12,k)
\right]
\\
&-
\frac{\Delta t}{\Delta z}
\left[
B^{\,n+\frac12}_y(i+\tfrac12,j,k+\tfrac12)
-
B^{\,n+\frac12}_y(i+\tfrac12,j,k-\tfrac12)
\right]
\\
&-
\Delta t\,
J^{\,n+\frac12}_x(i+\tfrac12,j,k).
\end{aligned}
\label{EQ_Yee_Ex}
\end{equation}

The remaining field components are updated analogously by cyclic permutation of the spatial coordinates and therefore are omitted for brevity.

As with any explicit FDTD scheme, the numerical stability of the Yee algorithm is constrained by the Courant--Friedrichs--Lewy (CFL) condition~\cite{Yee1966,Taflove2005}. For a three-dimensional Cartesian mesh, the time step must satisfy

\begin{equation}
c\Delta t
\le
\frac{1}
{\sqrt{
\frac1{(\Delta x)^2}
+
\frac1{(\Delta y)^2}
+
\frac1{(\Delta z)^2}
}},
\label{eq:CFL_general}
\end{equation}

which guarantees the stable propagation of electromagnetic information across the computational grid. In addition to satisfying the CFL criterion, the use of centered finite differences provides a second-order accurate discretization in both space and time.

The local nature of the Yee update equations makes the algorithm particularly well suited for both shared-memory and distributed-memory parallel implementations, since each field component depends only on neighboring grid values. This locality minimizes data dependencies, facilitates domain decomposition and parallel execution, and constitutes one of the principal reasons why the Yee scheme remains the field solver of choice in large-scale electromagnetic PIC simulations.
\subsection{Super-particle representation and particle integration}
Within the particle-in-cell approximation, the continuous distribution function is represented by a finite ensemble of super-particles (SPs), each corresponding to a large number of physical particles occupying a finite region of phase space~\cite{BirdsallLangdon2004,HockneyEastwood1988}. While the momentum distribution of every SP is approximated by a Dirac delta function, its spatial distribution is described by a finite-size shape function. Accordingly, the distribution associated with an individual SP can be written as

\begin{equation}
f_{sp}(t,\mathbf{r},\mathbf{p})
=
N_p\,
S(\mathbf{r}-\mathbf{r}_{sp})
\,
\delta(\mathbf{p}-\mathbf{p}_{sp}),
\label{EQ_SP_distribution}
\end{equation}

where $N_p$ denotes the number of physical particles represented by the SP, $S(\mathbf r-\mathbf r_{sp})$ is its spatial shape function, and $(\mathbf r_{sp},\mathbf p_{sp})$ are the corresponding position and momentum in phase space. Consequently, a super-particle should not be interpreted as a point charge but as a finite charged cloud whose spatial extent is determined by the adopted shape function~\cite{BirdsallLangdon2004,HockneyEastwood1988}, commonly constructed from B-spline basis functions of different orders. In three dimensions, the particle shape is expressed as

\begin{equation}
S(\mathbf r-\mathbf r_{sp})
=
\frac{1}{\Delta x\,\Delta y\,\Delta z}
\,b_l\!\left(\frac{x-x_{sp}}{\Delta x}\right)
\,b_l\!\left(\frac{y-y_{sp}}{\Delta y}\right)
\,b_l\!\left(\frac{z-z_{sp}}{\Delta z}\right),
\label{EQ_shape}
\end{equation}

where $b_l$ denotes the B-spline basis function of order $l$, while the normalization factor ensures that the shape function integrates to unity over the computational domain. The choice of $l$ defines the spatial representation of the super-particle and, consequently, determines the structure of the particle--mesh operators employed throughout the PIC algorithm. In the present implementation, zeroth-order B-splines are adopted, resulting in first-order (cloud-in-cell) field interpolation together with a charge-conserving current deposition scheme, as described in the following subsections. Consequently, particle-to-grid charge and current deposition and grid-to-particle electromagnetic-field interpolation constitute complementary numerical operations whose formulation is determined by the adopted particle shape function. This choice directly affects the accuracy, conservation properties, and susceptibility of the numerical scheme to grid-scale noise.

Within this framework, the electromagnetic fields experienced by each SP correspond to the weighted average of the grid fields over its spatial distribution,

\begin{equation}
\mathbf E_{sp}
=
\int
S(\mathbf r-\mathbf r_{sp})
\,
\mathbf E(\mathbf r)
\,d^3\mathbf r,
\qquad
\mathbf B_{sp}
=
\int
S(\mathbf r-\mathbf r_{sp})
\,
\mathbf B(\mathbf r)
\,d^3\mathbf r,
\label{EQ_particle_fields}
\end{equation}

thereby avoiding the singular field evaluation associated with point particles. The electromagnetic fields acting on each SP therefore arise naturally from its finite spatial extent rather than from a pointwise evaluation of the fields.

The dynamics of every SP are governed by the relativistic Newton--Lorentz equations introduced in Section~\ref{numerical_model}. Their numerical integration is performed using the Boris algorithm~\cite{Boris1970}, which remains one of the most widely used particle pushers in electromagnetic PIC simulations owing to its simplicity, second-order temporal accuracy, time reversibility, phase-space volume preservation, and favorable long-term energy behavior~\cite{BirdsallLangdon2004}. Furthermore, its leapfrog formulation naturally matches the staggered temporal discretization adopted by the Yee field solver.

Accordingly, particle momenta are defined at half-integer time levels, whereas particle positions are stored at integer instants,

\begin{align}
\frac{\mathbf p^{\,n+\frac12}_{sp}
-
\mathbf p^{\,n-\frac12}_{sp}}
{\Delta t}
&=
q_{sp}
\left(
\mathbf E^{\,n}_{sp}
+
\mathbf v^{\,n}_{sp}
\times
\mathbf B^{\,n}_{sp}
\right),
\\
\frac{\mathbf r^{\,n+1}_{sp}
-
\mathbf r^{\,n}_{sp}}
{\Delta t}
&=
\mathbf v^{\,n+\frac12}_{sp}.
\end{align}

Rather than solving the coupled Lorentz equation directly, the Boris algorithm treats the electric and magnetic effects separately through a sequence of explicit operations consisting of a half electric acceleration, a rotation in the magnetic field, and a second half electric acceleration. The particle position is subsequently advanced using the updated momentum. This operator splitting preserves the second-order temporal accuracy of the leapfrog scheme while avoiding the solution of implicit equations.

Because both the effective electromagnetic fields acting on the particles and the charge and current densities deposited onto the computational grid are determined by the adopted particle shape function, the interpolation and deposition procedures constitute two complementary aspects of the same particle--mesh coupling. Their numerical implementation is presented in the following subsections, where both operations are formulated consistently from the same particle shape function.
\subsection{Field interpolation and source deposition}
The particle--mesh coupling constitutes the interface between the Lagrangian description of the super-particles and the Eulerian representation of the electromagnetic fields. During each PIC iteration, two complementary transfer operations are required. First, the electromagnetic fields defined on the computational mesh are interpolated to the super-particle positions in order to evaluate the Lorentz force. Subsequently, the charge and current carried by the super-particles are projected back onto the mesh, providing the source terms required by Maxwell's equations. Both operations are derived consistently from the same particle shape function, thereby ensuring a self-consistent coupling between particles and fields throughout the simulation.

In the present implementation, each super-particle is represented by a zeroth-order B-spline, \(b_0\), corresponding to a uniform finite-size charged cloud. Following the standard PIC formulation, Eq.~(\ref{EQ_particle_fields}), the associated particle--mesh weighting functions are obtained from the first-order B-spline, \(b_1\)~\cite{BirdsallLangdon2004,HockneyEastwood1988}. Consequently, both field interpolation and source deposition reduce to the conventional cloud-in-cell (CIC) scheme, ensuring a consistent particle--mesh representation. The current density, however, is deposited using the charge-conserving ZigZag scheme described below, consistently with the same representation.

In KEMPIC-3D, field interpolation is performed directly on the staggered Yee lattice rather than on an auxiliary collocated mesh. Therefore, each component of $\mathbf{E}$ and $\mathbf{B}$ is interpolated from its own physical location within the Yee cell. This approach avoids an additional interpolation stage and provides a more consistent evaluation of the Lorentz force within the same discrete geometry employed by the electromagnetic field solver.

Since the magnetic field is naturally defined at half-integer time levels in the Yee scheme, whereas the electric field and particle positions are evaluated at integer time levels during the force calculation, the magnetic field required by the Boris pusher is reconstructed at \(t^n\) through temporal centering,

\begin{equation}
\mathbf{B}^{\,n}
=
\frac{
\mathbf{B}^{\,n+\frac{1}{2}}
+
\mathbf{B}^{\,n-\frac{1}{2}}
}{2},
\label{EQ_B_centered}
\end{equation}

which preserves the second-order temporal accuracy of the leapfrog integration. The resulting magnetic field is then interpolated to the super-particle position using the same CIC weights employed for the electric field.

After particle advancement, the source terms are deposited back onto the computational mesh using the same particle representation. The charge density is assigned using the standard CIC weighting procedure. The current density, however, requires a charge-conserving deposition scheme because inconsistencies between $\rho$ and $\mathbf{J}$ may lead to violations of the discrete continuity equation, Eq.~(\ref{eq_continuity_equation}). Numerous charge-conserving deposition algorithms have been proposed for
electromagnetic PIC simulations, including the general scheme of
Esirkepov~\cite{Esirkepov2001}, the computationally efficient ZigZag
algorithm of Umeda \textit{et al.}~\cite{Umeda2003}, and the more recent
EZ algorithm of Steiniger \textit{et al.}~\cite{EZ}, which
hybridizes the Esirkepov and ZigZag approaches to improve computational
performance while retaining exact charge conservation. In the present implementation, current deposition is performed using the original ZigZag algorithm of Umeda \textit{et al.}, which is particularly well suited to the particle representation adopted here and deposits the current components directly at the staggered locations required by the Yee discretization.

Because the implementation follows the original formulation of Umeda \textit{et al.}, its complete derivation is not reproduced here. For the purposes of the present implementation, the essential property is that the deposited current satisfies the discrete continuity equation exactly and can therefore be directly incorporated into the discretized Ampère--Maxwell equation without additional interpolation or correction procedures. Consequently, field interpolation and source deposition remain fully consistent particle--mesh transfer operations derived from the same finite-size representation of the super-particles~\cite{Umeda2003}.

The numerical components described throughout this section are assembled into the computational workflow summarized in Fig.~\ref{FIG_PIC_algorithm}. The diagram illustrates how the field solver, particle pusher, field interpolation, and charge-conserving current deposition are integrated into a single self-consistent particle--field cycle. Starting from the electromagnetic fields and particle quantities defined at the appropriate staggered time levels, the magnetic field is first advanced using Faraday's law. A temporally centered magnetic field is then reconstructed and used, together with the electric field, to advance the particle trajectories through the Boris algorithm. The updated particle trajectories are subsequently used to deposit the charge-conserving current density onto the Yee mesh using the ZigZag scheme, after which the electric field is advanced through the discretized Ampère--Maxwell equation. The updated electromagnetic fields and particle quantities then provide the input for the next iteration, completing one PIC time step.

This workflow highlights the close coupling between the Lagrangian particle description and the Eulerian electromagnetic field solver that characterizes fully electromagnetic PIC simulations. Additional numerical modules, including boundary-condition treatment, current filtering, moving-window operation, particle initialization, and external electromagnetic sources, are integrated into the overall computational framework without altering the fundamental structure of the particle--field coupling.
\subsection{Computational architecture, accessibility, and extensibility}
\label{sec:computational_architecture}
KEMPIC-3D has been developed with particular emphasis on direct access to the numerical procedures that constitute the PIC implementation. The principal computational operations, including electromagnetic field evolution, particle integration, particle--field interpolation, charge and current deposition, boundary-condition treatment, particle initialization, external electromagnetic sources, diagnostics, and moving-window operation, are organized into identifiable numerical modules. This organization allows the user to inspect and modify the routines associated with the main stages of the simulation.

A central design principle of KEMPIC-3D is algorithmic transparency. Rather than encapsulating the numerical procedures within highly abstract computational layers, the implementation preserves a direct correspondence between the algorithms described in the preceding sections and their realization in the source code. This facilitates the examination of the numerical methods and the modification or replacement of individual components when required.

The modular organization also provides a basis for the extensibility of the code. Additional physical models, numerical algorithms, and diagnostic capabilities can be incorporated without altering the fundamental structure of the self-consistent particle--field coupling. In this sense, KEMPIC-3D is intended not only as a simulation code for the applications considered here but also as a computational framework that can be adapted to different problems in kinetic plasma physics.

The present implementation incorporates a lightweight parallelization strategy to reduce computational time in multidimensional simulations. Its primary design objective, however, is not large-scale high-performance computing, but rather numerical transparency, versatility, and ease of modification. KEMPIC-3D therefore complements, rather than seeks to reproduce the computational philosophy of, established PIC platforms specifically optimized for massively parallel and large-scale simulations.

To promote reproducibility and facilitate independent use and further development, the KEMPIC-3D source code will be made publicly available through a GitHub repository. The repository will provide access to the numerical implementation together with representative input files and documentation, enabling users to run, inspect, and modify the code. The numerical verification and benchmarking of the KEMPIC-3D implementation are presented in the following section.
\section{Code verification and numerical benchmarking}
\label{sec:verification}
The purpose of this section is to assess the numerical accuracy, stability, and self-consistency of KEMPIC-3D in describing the kinetic and electromagnetic dynamics governed by the Vlasov--Maxwell system. To this end, the code is subjected to a systematic verification and numerical benchmarking procedure aimed at assessing both the correctness of the implemented numerical algorithms and the ability of the complete self-consistent PIC model to reproduce well-established behavior in representative plasma benchmark problems.

In the context of computational plasma physics, code verification and numerical benchmarking address complementary but distinct objectives. Code verification determines whether the numerical algorithms correctly represent the governing mathematical equations and whether the individual components of the implementation perform as intended. Numerical benchmarking, on the other hand, assesses whether the coupled particle--field model reproduces the theoretically established physical behavior of representative benchmark problems~\cite{Roache1998,Oberkampf2010}. Together, these procedures provide confidence in both the numerical consistency of the implementation and the physical reliability of the resulting simulations.

Following this approach, the assessment strategy adopted here is organized hierarchically. The first stage independently verifies the principal numerical modules, thereby allowing their individual performance to be isolated and evaluated. The relativistic Boris particle pusher is verified through comparison with analytical charged-particle trajectories in prescribed electromagnetic fields. The Yee electromagnetic field solver is subsequently assessed through the propagation of the fundamental TE\(_{10}\) mode in a rectangular waveguide, together with a grid-convergence analysis.

After verification of these fundamental components, the self-consistent particle--field coupling is assessed through a cold plasma oscillation benchmark. The resulting collective plasma response is compared with the theoretical plasma frequency, while the conservation properties and preservation of the discrete continuity equation are also evaluated. Taken together, these tests evaluate the accuracy, stability, conservation properties, and self-consistency of the KEMPIC-3D implementation.
\subsection{Verification of the Boris particle pusher}
The first stage of the code verification assesses the numerical accuracy of the relativistic Boris particle pusher independently of the electromagnetic field solver. Since the Boris algorithm advances particle trajectories by treating the electric and magnetic contributions of the Lorentz force separately, two complementary analytical benchmarks were considered. The first isolates the magnetic rotation by prescribing a uniform magnetic field in the absence of an electric field, whereas the second examines the electric acceleration under a uniform electric field with no magnetic field. Together, these benchmarks provide verification of the electric-acceleration and magnetic-rotation operations underlying the Boris integration scheme.

As a first benchmark, the trajectory of a single electron was computed in the presence of a prescribed uniform magnetic field $\mathbf{B}=1\,\hat{z}$ in normalized units, and in the absence of an electric field ($\mathbf{E}=0$). Under these conditions, the analytical solution corresponds to uniform relativistic cyclotron motion with constant kinetic energy and a fixed Larmor radius, making it an ideal test for evaluating the temporal accuracy and long-term stability of the magnetic rotation performed by the Boris algorithm.

The electron was initialized with a velocity $v_y/c=0.99$, perpendicular to the magnetic field, corresponding to a Lorentz factor \(\gamma \approx7.09\). The temporal resolution was systematically refined using $11$, $21$, $41$, $81$, $161$, $321$, $641$, and $1281$ points per cyclotron period. For each case, the time step was defined as \( \Delta t=\frac{2\pi\gamma}{N_{\mathrm{pt}}-1},\) where $N_{\mathrm{pt}}$ denotes the number of temporal samples over one cyclotron period and $\gamma$ is the relativistic Lorentz factor. In the adopted normalized units, the corresponding relativistic cyclotron period is \(T_c = 2\pi\gamma\). To assess the long-term behavior of the integrator, every simulation was propagated over $1000$ complete cyclotron periods.

\begin{figure}[t]
    \centering
    \includegraphics[width=\linewidth]{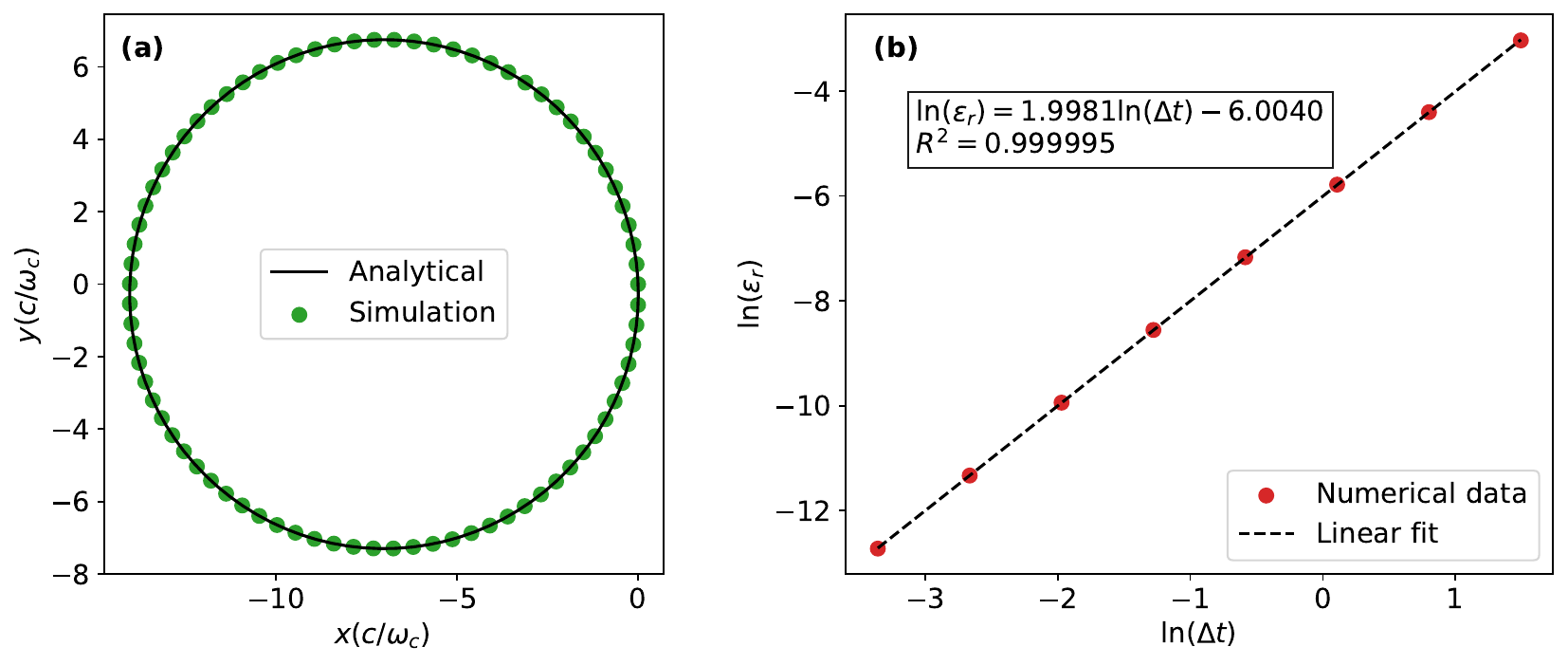}
    \caption{Verification benchmark of the magnetic rotation stage of the relativistic Boris particle pusher. (a) Comparison between the numerical and analytical cyclotron trajectories for an electron moving in a uniform magnetic field ($\mathbf{B}=1\,\hat{z}$, normalized units) with $\mathbf{E}=0$. (b) Log--log convergence of the relative Larmor-radius error as a function of the time step. The fitted slope of $1.998$ confirms the expected second-order temporal accuracy of the Boris integration scheme.}
    \label{FIG_Boris_convergence}
\end{figure}

Figure~\ref{FIG_Boris_convergence} summarizes the results obtained for the magnetic benchmark. Panel (a) compares the numerical and analytical trajectories over the final cyclotron revolution after $1000$ simulated gyrations. The nearly perfect overlap between the two solutions demonstrates the absence of noticeable orbit distortion or secular numerical drift, highlighting the excellent long-term stability of the Boris particle pusher. Such stability is essential for fully electromagnetic PIC simulations, in which particles are routinely advanced over millions of time steps while maintaining accurate trajectories.

Panel (b) presents the log--log convergence of the relative Larmor-radius error, $\varepsilon_r$, as a function of the time step $\Delta t$. The numerical Larmor radius obtained for each temporal resolution was compared with the analytical solution, and the resulting error was fitted using a least-squares linear regression. The fitted slope is $1.998$, with a coefficient of determination $R^2=0.999995$, yielding \( \varepsilon_r \propto (\Delta t)^{1.998},\) in excellent agreement with the theoretical second-order temporal accuracy of the Boris integrator. The near-unity coefficient of determination indicates that the measured error is dominated by the expected truncation error of the numerical scheme rather than by accumulated round-off errors or numerical instabilities.

It is worth noting that, for the purely magnetic benchmark, the particle speed, kinetic energy, and Lorentz factor remained constant throughout the simulation within double-precision round-off accuracy, exhibiting relative variations of the order of $10^{-15}$. This further confirms the excellent conservation properties of the implemented Boris integrator.

The second benchmark independently assesses the electric component of the Lorentz force by considering the relativistic acceleration of a single electron in a prescribed uniform electric field. In this case, the magnetic field was set to zero ($\mathbf{B}=0$), while a constant normalized electric field $\mathbf{E}=1\,\hat{x}$ was applied. The electron was initially released from rest at the origin, and its subsequent motion was followed over the normalized time interval $0\leq t\leq1$, for which an analytical relativistic solution exists.

The temporal resolution was systematically refined using $101$, $201$, $401$, $801$, $1601$, $3201$, $6401$, and $12801$ uniformly distributed time samples over the integration interval. For each simulation, the numerical particle position at the final time was compared with the analytical solution, and the relative position error along the direction of acceleration, $\varepsilon_x$, was evaluated.

\begin{figure}[t]
\centering
\includegraphics[width=\linewidth]{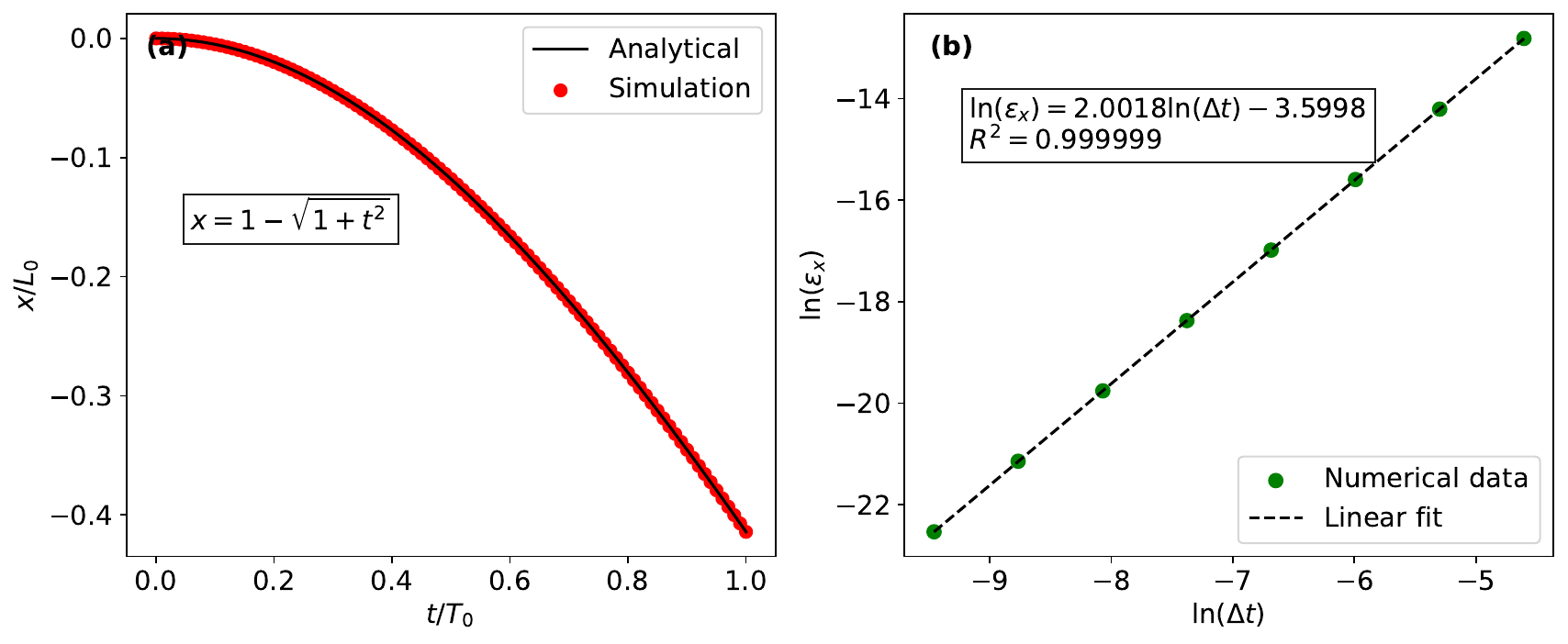}
\caption{Verification benchmark of the electric acceleration stage of the relativistic Boris particle pusher. (a) Comparison between the numerical and analytical evolution of the position $x(t)$ of an electron initially at rest and accelerated by a uniform electric field ($\mathbf{E}=1\,\hat{x}$, normalized units), with $\mathbf{B}=0$. (b) Log--log convergence of the relative position error as a function of the time step. The fitted slope of approximately $2.002$ confirms the expected second-order temporal accuracy of the Boris integration scheme.}
\label{FIG_Boris_Electric}
\end{figure}

Figure~\ref{FIG_Boris_Electric} summarizes the verification results for the electric benchmark. Panel (a) compares the numerical and analytical evolution of the particle position over the normalized time interval $0\leq t\leq1$. The nearly perfect overlap between both solutions demonstrates that the Boris algorithm accurately reproduces the relativistic electric acceleration predicted by the analytical solution. Panel (b) presents the corresponding log--log convergence of the relative position error as a function of the time step. The fitted slope is $2.002$, with a coefficient of determination $R^2=0.999999$, yielding \( \varepsilon_x \propto (\Delta t)^{2.002},\) again confirming the expected second-order temporal accuracy of the Boris integration scheme.

Although the convergence analysis is based on the particle position, the relative error in the particle velocity at the final simulation time remains close to double-precision round-off accuracy for all temporal resolutions considered, ranging from approximately $10^{-16}$ to $10^{-13}$. The slight increase observed for the smallest time steps is consistent with the accumulation of floating-point round-off errors as the number of integration steps increases, indicating that the temporal discretization error has become negligible compared with the finite precision of double-precision arithmetic.

Taken together, the two benchmarks show excellent agreement with the corresponding analytical solutions and recover the expected second-order temporal convergence of the Boris scheme. These results confirm the numerical accuracy and robustness of the particle-pushing implementation adopted in KEMPIC-3D, while the magnetic-rotation benchmark additionally demonstrates its long-time numerical stability.
\subsection{Verification of the electromagnetic field solver}
The second stage of the code verification assesses the accuracy of the implementation of the Yee finite-difference time-domain (FDTD) solver independently of the particle dynamics. For this purpose, the propagation of the fundamental $\mathrm{TE}_{10}$ mode inside a perfectly conducting rectangular waveguide was considered, since an analytical solution is available for both its temporal evolution and spatial field distribution.

The waveguide dimensions were selected as $a=3~\mathrm{cm}$, $b=a/2$, and $L_z=1~\mathrm{m}$. For these dimensions, the cut-off frequency of the fundamental mode is $f_c^{TE_{10}}=c/(2a)\approx5~\text{GHz}$. A monochromatic excitation at $f_0=9~\mathrm{GHz}>f_c^{TE_{10}}$ was prescribed at the input plane $z=0$. The excitation frequency is adopted as the normalization parameter for time, such that the normalized time unit corresponds to one excitation period $T_0=1/f_0$.

To minimize startup transients, the sinusoidal excitation was gradually introduced using the exponential envelope \( g(t)=1-\exp\!\left(-\frac{t}{\tau}\right),\) with $\tau=2$, corresponding to two excitation periods. This smooth temporal modulation suppresses the high-frequency components associated with an abrupt source turn-on and reduces spurious numerical transients, allowing the guided mode to evolve smoothly toward a steady-state monochromatic regime.

\begin{figure}[t]
\centering
\includegraphics[width=\linewidth]{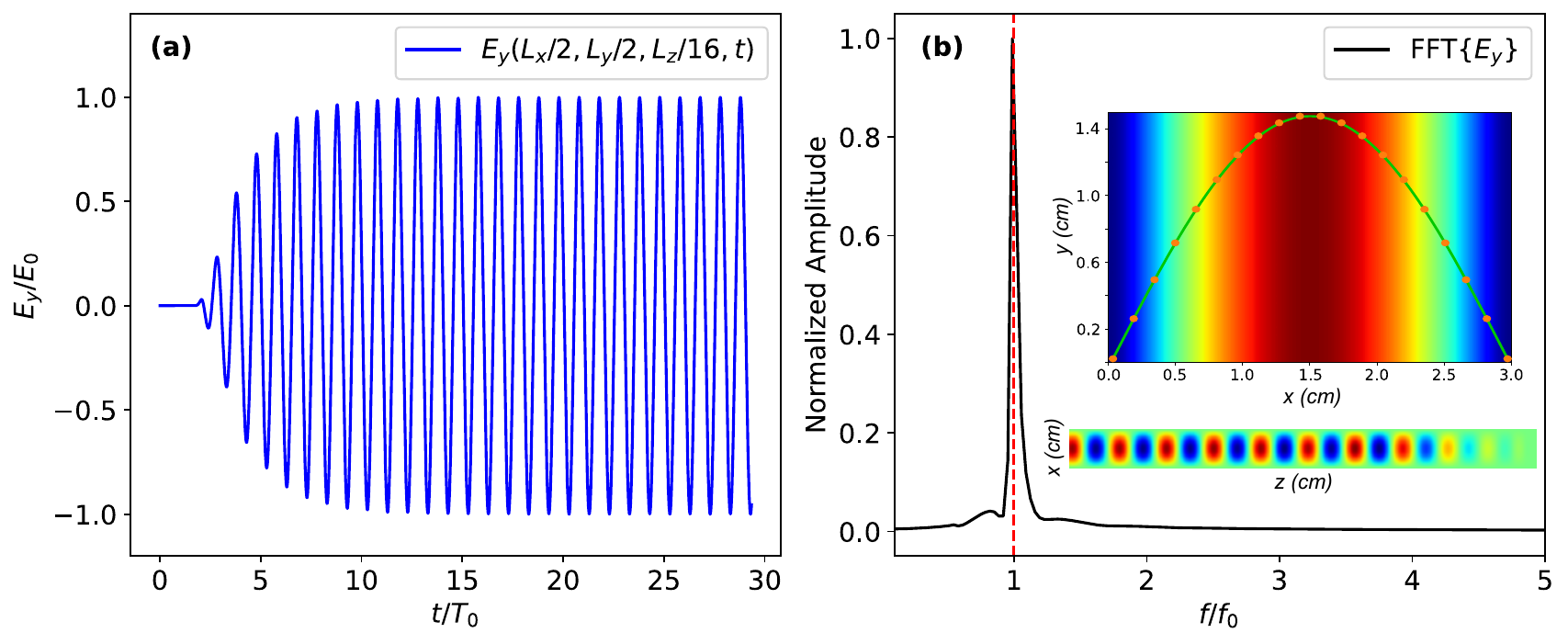}
\caption{Verification benchmark of the Yee FDTD solver using propagation of the fundamental $\mathrm{TE}_{10}$ mode in a perfectly conducting rectangular waveguide. (a) Temporal evolution of the normalized electric field component $E_y$ measured at $(L_x/2,L_y/2,L_z/16)$. (b) Normalized Fourier spectrum of the measured signal. The insets show the transverse electric-field profile across the waveguide cross section and the two-dimensional field distribution in the longitudinal plane, both reproducing the analytical spatial structure of the $\mathrm{TE}_{10}$ mode.}
\label{FIG_Yee_Verification}
\end{figure}

Figure~\ref{FIG_Yee_Verification} summarizes the verification benchmark. Panel (a) presents the temporal evolution of the electric field component $E_y$ measured at $(L_x/2,L_y/2,L_z/16)$. The progressive increase of the field amplitude demonstrates the smooth establishment of the driven mode, followed by a stable steady-state oscillation at the prescribed excitation frequency.

Panel (b) shows the normalized Fourier spectrum of the same signal. A single dominant peak appears at the normalized excitation frequency $f=1$, confirming that the numerical field oscillates at the prescribed excitation frequency without significant spurious spectral components.

The insets further verify the spatial characteristics of the numerical solution. The transverse electric-field profile across the waveguide cross section and the two-dimensional field distribution accurately reproduce the analytical structure of the fundamental $\mathrm{TE}_{10}$ mode. These results demonstrate that the implemented Yee FDTD solver correctly captures the temporal and spatial properties of the guided electromagnetic field.

A grid-convergence study was subsequently performed by systematically refining the computational mesh while maintaining approximately uniform resolution in the three spatial directions, $\Delta x\approx\Delta y\approx\Delta z$. For each mesh, the time step was selected according to the Courant stability criterion, maintaining a Courant number of $0.5$. The resulting numerical field amplitude was compared with the corresponding analytical solution, and the relative errors are summarized in Table~\ref{TAB_Yee_Convergence}. A characteristic grid spacing was defined as \( \Delta h=\sqrt{(\Delta x)^2+(\Delta y)^2+(\Delta z)^2},\) which corresponds to the diagonal length of the three-dimensional computational cell and provides a characteristic measure of the grid spacing during the refinement process.

The observed convergence rate of $2.136$, with a coefficient of determination $R^2=0.972$, is consistent with the expected second-order accuracy of the Yee FDTD scheme. The slight deviations from the ideal quadratic behavior are attributed to the simultaneous refinement of both the spatial and temporal discretizations imposed by the Courant stability condition.

\begin{table}[t]
\centering
\caption{Grid-convergence study of the Yee FDTD solver for propagation of the fundamental $\mathrm{TE}_{10}$ mode. Here, $\Delta h=\sqrt{(\Delta x)^2+(\Delta y)^2+(\Delta z)^2}$ denotes the characteristic spatial discretization, and $\varepsilon_{E_0}$ is the relative error of the steady-state electric-field amplitude with respect to the analytical solution.}
\label{TAB_Yee_Convergence}
\begin{tabular}{cccc}
\hline
$n_x\times n_y\times n_z$ &
$\Delta h$ &
$n_t$ &
$\varepsilon_{E_0}$ \\
\hline
$12\times6\times376$     & $4.853\times10^{-3}$ & $1251$  & $4.54\times10^{-2}$ \\
$23\times12\times751$    & $2.345\times10^{-3}$ & $2501$  & $4.37\times10^{-3}$ \\
$46\times23\times1501$   & $1.164\times10^{-3}$ & $5001$  & $2.95\times10^{-3}$ \\
$91\times46\times3001$   & $5.774\times10^{-4}$ & $10001$ & $3.74\times10^{-4}$ \\
$181\times91\times6001$  & $2.887\times10^{-4}$ & $20001$ & $8.41\times10^{-5}$ \\
\hline
\end{tabular}
\end{table}

These results demonstrate that the implemented Yee FDTD solver accurately reproduces both the temporal and spatial characteristics of the guided electromagnetic mode considered here. The observed convergence rate is also consistent with the expected second-order accuracy of the Yee scheme under the coupled space-time refinement, supporting the numerical consistency of the electromagnetic field solver implemented in KEMPIC-3D.
\subsection{Verification of the self-consistent particle--field coupling}
After independently verifying the particle pusher and the electromagnetic field solver, the next stage assesses the self-consistent evolution of the coupled particle--field system. Unlike the previous benchmarks, in which each numerical module was examined separately, the present test evaluates their coupled operation throughout the simulation. A linear cold-plasma oscillation is used as the benchmark because its characteristic frequency is analytically known, allowing the collective response of the coupled numerical system to be directly compared with theory.

A homogeneous cold plasma equilibrium was first initialized. The plasma distribution was initialized using a quiet-start particle loading in order to minimize statistical fluctuations and isolate the collective plasma dynamics. A small longitudinal perturbation was then imposed, establishing an effectively one-dimensional electrostatic configuration along the $z$-direction while retaining the full three-dimensional electromagnetic formulation of the code.

The computational domain was defined as $L_x=L_y=L_z=2\pi$ in normalized units. The transverse directions were discretized using only two grid points ($N_x=N_y=2$) with all quantities initialized uniformly in the transverse directions and periodic boundary conditions imposed so that the dynamics remained effectively one-dimensional. The longitudinal direction was discretized using $N_z=101$ grid points, corresponding to one wavelength of the imposed $k=1$ fundamental mode. Periodic boundary conditions were applied along all three directions, reproducing an effectively infinite homogeneous plasma. The time step was determined from the Courant stability criterion using a Courant number of $0.5$.

A small sinusoidal perturbation was imposed on the longitudinal particle velocity, \( v_z(z,t=0)=10^{-3}\sin(z),\) while the remaining velocity components were initialized to zero. The perturbation amplitude was sufficiently small to keep the plasma within the linear regime, for which the analytical solution predicts harmonic oscillations at the electron plasma frequency. Consequently, deviations from the expected linear response primarily reflect numerical discretization and particle--field coupling rather than nonlinear plasma effects.

\begin{figure}[t]
\centering
\includegraphics[width=\linewidth]{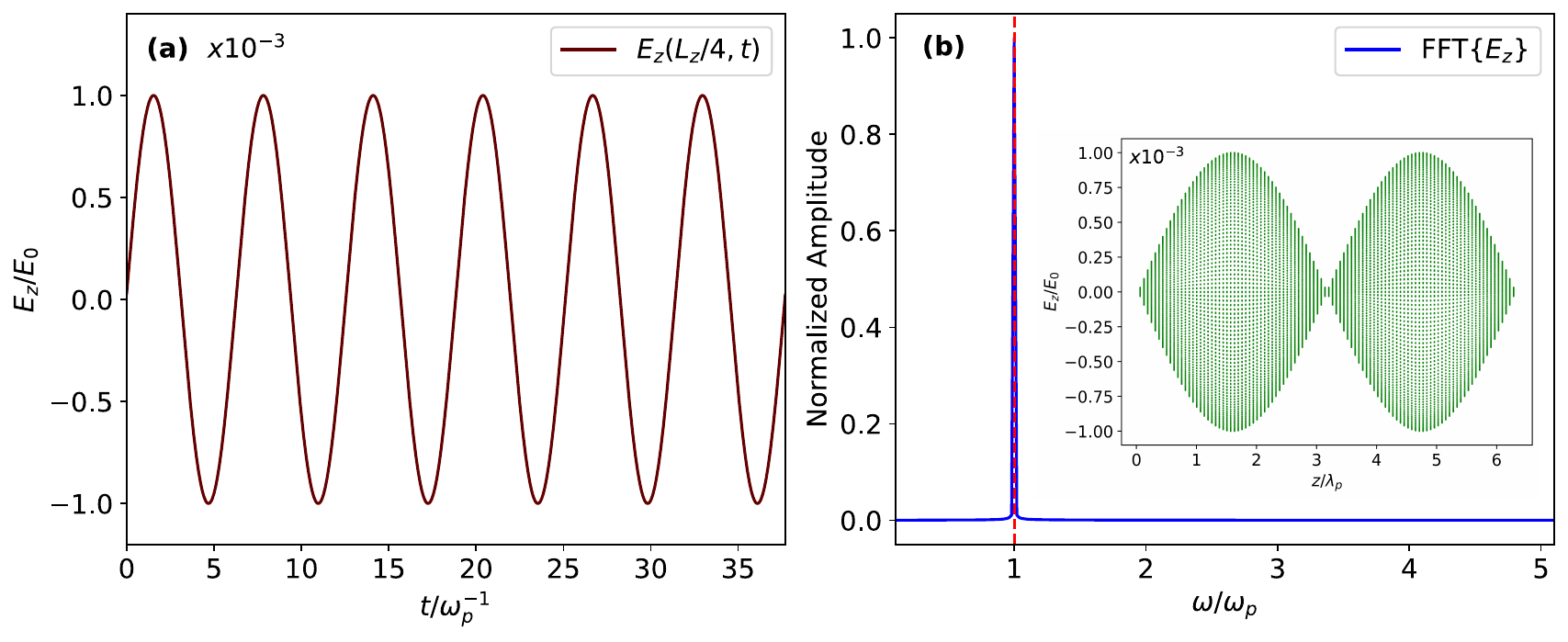}
\caption{Verification benchmark of the fully coupled electromagnetic PIC algorithm using plasma oscillations. (a) Temporal evolution of the longitudinal electric field component $E_z$ measured at $z=L_z/4$. (b) Normalized Fourier spectrum of the same signal. The inset shows snapshots of the longitudinal electric field illustrating the standing electrostatic wave supported by the periodic domain.}
\label{FIG_Plasma_Oscillations}
\end{figure}

Figure~\ref{FIG_Plasma_Oscillations} summarizes the results of the plasma-oscillation benchmark. Panel (a) presents the temporal evolution of the longitudinal electric field component $E_z$ measured at $z=L_z/4$. Following the initial perturbation, the plasma exhibits sustained oscillations with nearly constant amplitude, indicating the absence of significant numerical damping or growth over the simulated interval.

Panel (b) shows the normalized Fourier spectrum of the same signal. A single narrow spectral peak is observed at $\omega/\omega_p = 1$, in excellent agreement with the theoretical prediction. The inset presents snapshots of the longitudinal electric field along the propagation direction, illustrating the standing electrostatic wave associated with the fundamental mode supported by the periodic domain.

This benchmark provides a direct assessment of the self-consistent particle--field coupling in KEMPIC-3D. The agreement between the simulated and theoretical plasma frequencies confirms that the coupled numerical system correctly reproduces the characteristic collective response of a cold plasma in the linear regime.

\begin{figure}[t]
\centering
\includegraphics[width=\linewidth]{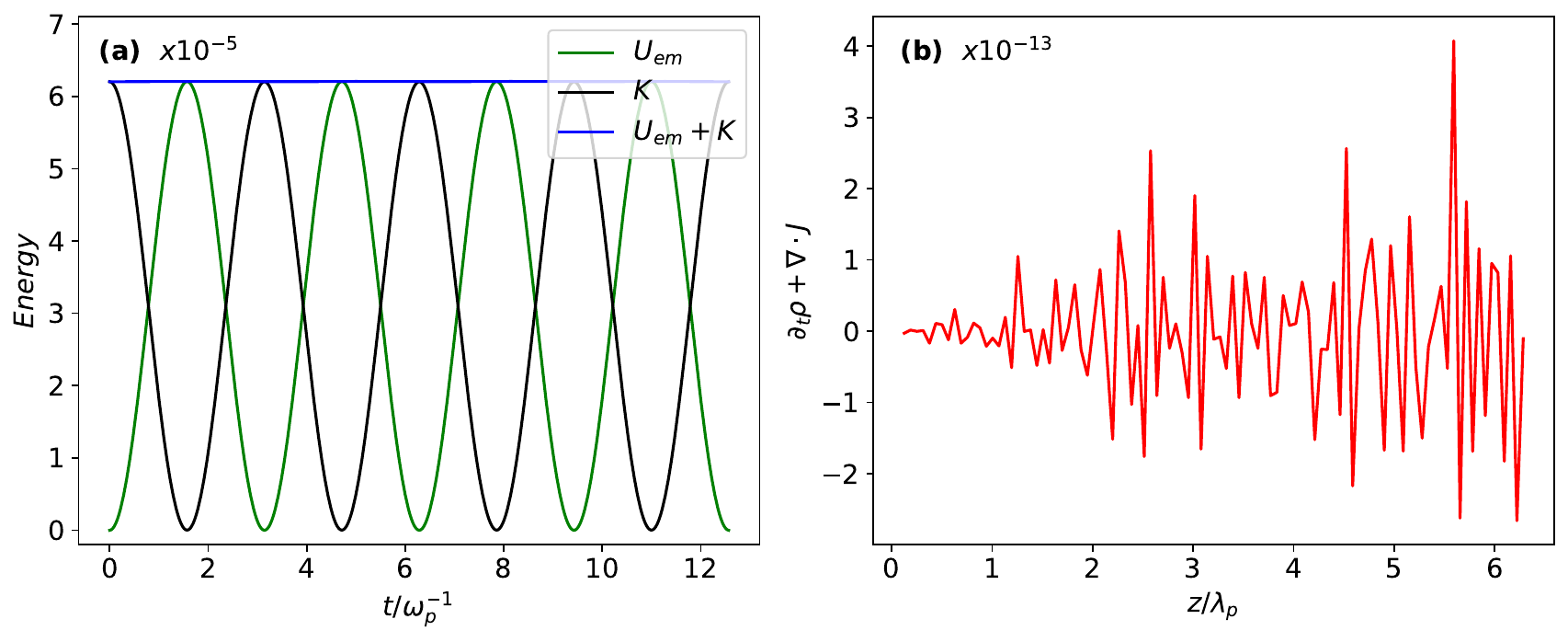}
\caption{Verification benchmark of the conservation properties of the fully coupled electromagnetic PIC algorithm. (a) Temporal evolution of the electromagnetic energy, particle kinetic energy, and total energy during the plasma-oscillation benchmark. (b) Residual of the discrete continuity equation evaluated along the longitudinal direction at the final simulation time.}
\label{FIG_PIC_Conservation}
\end{figure}

To further assess the self-consistency of the coupled particle--field dynamics, the conservation properties of the complete PIC implementation were examined. Figure~\ref{FIG_PIC_Conservation} summarizes the corresponding results. Panel (a) presents the temporal evolution of the electromagnetic energy, $U_{\mathrm{EM}}$, the particle kinetic energy, $K$, and their sum, $U_{\mathrm{tot}}=U_{\mathrm{EM}}+K$. As expected for plasma oscillations, the electromagnetic and particle kinetic energies oscillate out of phase, while the total energy remains nearly constant throughout the simulation. The relative variation of the total energy remains below $7.5\times10^{-4}$ over the entire simulation, demonstrating global energy conservation at this level of numerical accuracy.

Panel (b) evaluates the discrete continuity equation computed along the longitudinal direction at the final simulation time. The residual remains at the level of $10^{-13}$ throughout the computational domain, close to double-precision round-off accuracy. This result demonstrates that the implemented charge-conserving current deposition preserves the discrete continuity equation to near machine precision, providing numerical consistency between charge evolution, current deposition, and electromagnetic field evolution.

Taken together, the plasma-oscillation benchmark and the conservation tests provide a comprehensive assessment of the self-consistent PIC implementation under linear collective plasma conditions. The accurate recovery of the theoretical plasma frequency, the preservation of total energy and discrete charge continuity, confirm the numerical accuracy, conservation properties, and robustness self-consistency of the KEMPIC-3D particle--field coupling.
\section{Representative application: Nonlinear laser-driven plasma wakefield generation}
Having verified the numerical implementation through a hierarchy of benchmark problems, the computational capabilities of KEMPIC-3D are finally demonstrated through a representative simulation of laser-driven plasma wakefield generation in an underdense plasma. The objective of this example is not to investigate electron injection or acceleration, but rather to illustrate the ability of the code to self-consistently reproduce the nonlinear laser--plasma interaction responsible for the excitation and evolution of plasma wakefields.
\subsection{Simulation setup}
Although KEMPIC-3D is fully three-dimensional, the present demonstration employs a reduced two-dimensional Cartesian configuration to reduce the computational cost while retaining the essential physics of laser-driven plasma wakefield generation. The computational domain is defined in the $x$--$z$ plane, where the laser pulse propagates along the longitudinal ($z$) direction and is linearly polarized along $y$. The third spatial dimension is represented by only two grid cells ($N_y=2$) together with periodic boundary conditions, thereby enforcing translational invariance along that coordinate. This reduced-dimensionality configuration is adopted exclusively for computational efficiency; the numerical algorithms and implementation remain identical to those employed in fully three-dimensional simulations.

An ultra-short Gaussian laser pulse propagates through an initially homogeneous underdense plasma. The ponderomotive force associated with the laser pulse drives a nonlinear plasma wake characterized by alternating regions of electron depletion and accumulation behind the laser pulse. Since no externally injected electron bunch is considered, the present example focuses exclusively on the self-consistent generation and evolution of the plasma wakefield.

The principal simulation parameters are summarized in Table~\ref{TAB_LWFA_parameters}. The selected values are representative of typical nonlinear laser--plasma interaction conditions and are intended primarily to demonstrate the capability of KEMPIC-3D to model laser-driven plasma wakefield generation.

\begin{table}[t]
\centering
\caption{Simulation parameters employed in the representative laser-driven plasma wakefield generation example.}
\label{TAB_LWFA_parameters}
\begin{tabular}{lc}
\hline
Parameter & Value \\
\hline
Plasma density, $n_0$ & $3.0\times10^{18}\,\mathrm{cm^{-3}}$ \\
Laser wavelength, $\lambda_0$ & $0.8\,\mu$m \\
Pulse duration (FWHM) & $28\,\mathrm{fs}$ \\
Laser waist, $w_0$ & $12\,\mu$m \\
Peak normalized vector potential, $a_0$ & 3.0 \\
Simulation window: $(L_x\times L_y\times L_z)$ & $(3\times1\times3)\lambda_p$ \\
Grid $(N_x\times N_y\times N_z)$ & $(301\times2\times1401)$ \\
Time step & Courant limited ($C=0.5$) \\
Moving window & Yes \\
\hline
\end{tabular}
\end{table}
\subsection{Wakefield evolution}
\begin{figure}[t]
\centering
\includegraphics[width=1.0\linewidth]{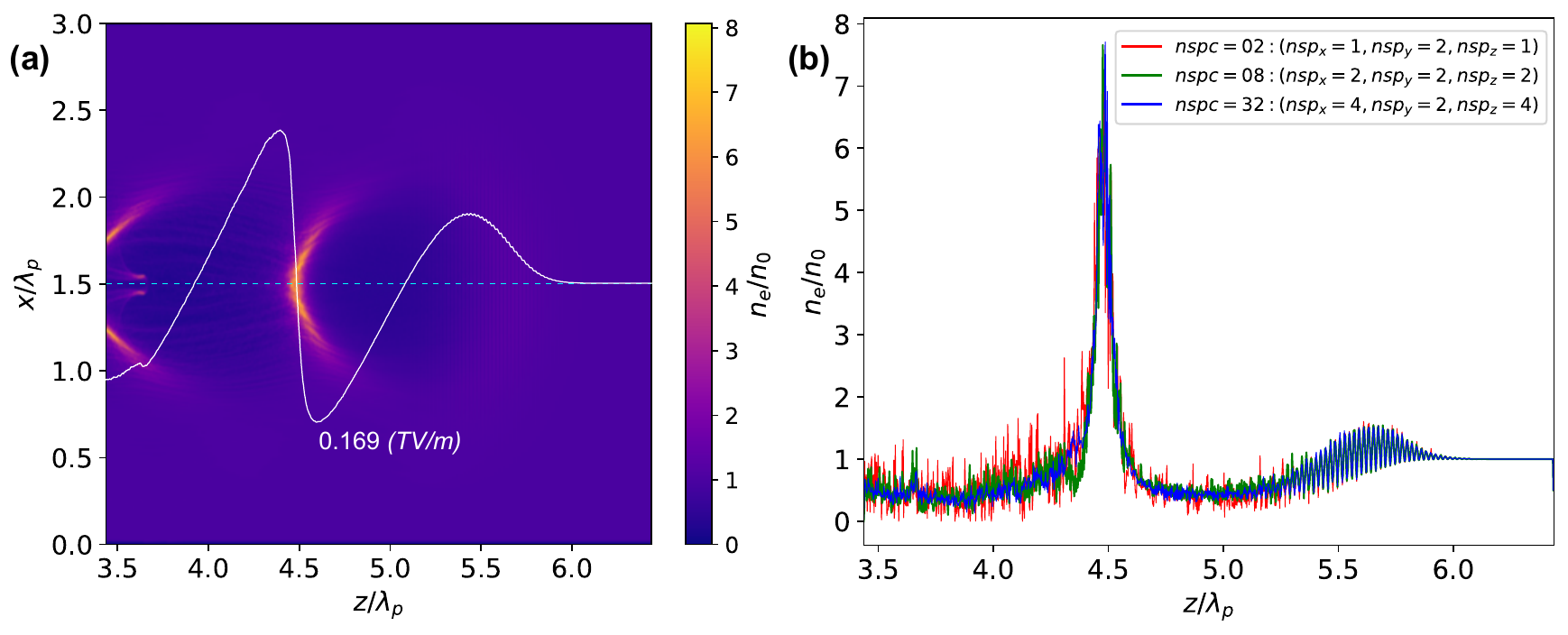}
\caption{Representative simulation of laser-driven plasma wakefield generation. (a) Normalized electron density distribution, $n_e/n_0$, at the end of the simulation. The white curve represents the longitudinal electric field, $E_z$, evaluated along the laser propagation axis. (b) Longitudinal electron density profiles extracted along the propagation axis for simulations performed using 2 (red), 8 (green), and 32 (blue) super-particles per cell.}
\label{FIG_LWFA}
\end{figure}

Figure~\ref{FIG_LWFA} presents representative results obtained with KEMPIC-3D for the self-consistent generation of a nonlinear plasma wakefield. The simulation was performed using 32 super-particles per cell, corresponding to four particles along the $x$ direction, two along $y$, and four along the $z$ direction. Panel (a) shows the normalized electron density distribution, $n_e/n_0$, at the end of the simulation, after the laser pulse has propagated approximately six plasma wavelengths, corresponding to about two moving-window lengths. The white curve superimposed on the density map represents the longitudinal accelerating electric field, $E_z$, evaluated along the laser propagation axis. The simulation reproduces the characteristic features of the nonlinear wakefield regime, including the formation of an electron density cavity followed by alternating regions of electron depletion and accumulation. The corresponding longitudinal electric field reaches a peak accelerating gradient of approximately $0.169~\mathrm{TV/m}$, demonstrating the capability of KEMPIC-3D to self-consistently capture the laser--plasma interaction responsible for wakefield generation.

To assess the influence of particle statistics on the numerical solution, panel (b) compares the longitudinal electron density profiles extracted along the propagation axis for simulations performed using 2 (red), 8 (green), and 32 (blue) super-particles per cell. These configurations correspond to particle distributions of $(1,2,1)$, $(2,2,2)$, and $(4,2,4)$ along the $(x,y,z)$ directions, respectively. As expected for particle-in-cell methods, increasing the number of super-particles reduces the statistical noise in the deposited charge density while preserving the overall wakefield structure. Despite a sixteen-fold increase in the number of super-particles per cell, all three simulations produce similar overall plasma-wave profiles, differing primarily in the level of statistical fluctuations.

No spatial filtering or smoothing techniques were applied to either the charge or current densities in these simulations. Although the deposited particle moments exhibit the statistical fluctuations characteristic of PIC methods, the computed electromagnetic fields remain comparatively smooth in the simulations considered. The principal wakefield features are preserved across the different particle numbers without the use of explicit spatial filtering or smoothing. These results demonstrate that KEMPIC-3D accurately reproduces the self-consistent formation of nonlinear plasma wakefields while maintaining robust numerical behavior without requiring additional filtering or smoothing procedures.

\subsection{Discussion}
The representative application presented in this section demonstrates the capability of KEMPIC-3D to self-consistently model the interaction between an intense laser pulse and an underdense plasma, leading to the excitation of a nonlinear plasma wakefield. The simulated electron density distributions and accelerating electric fields reproduce the characteristic features expected for the nonlinear wakefield regime, illustrating the ability of the code to capture the essential physics of multidimensional laser--plasma interactions.

The comparison performed using different numbers of super-particles per cell provides an additional assessment of the sensitivity of the numerical solution to particle statistics. Although increasing the particle number significantly reduces the statistical noise inherent to particle-in-cell methods, the overall wakefield structure and the resulting electromagnetic fields remain nearly unchanged, indicating that the principal features of the simulated plasma dynamics are not strongly dependent on particle discretization over the range considered in this work. Furthermore, these results were obtained without applying any spatial filtering or smoothing to the deposited charge or current densities, highlighting the stability of KEMPIC-3D under the conditions considered.

Although the purpose of this example is not to investigate laser wakefield acceleration itself, it provides a representative demonstration of the capability of KEMPIC-3D to model multidimensional laser--plasma interactions. Together with the systematic verification presented in the previous section, these results demonstrate the ability of the code to reproduce self-consistent electromagnetic plasma dynamics and support its application to future studies of laser-driven plasma acceleration and other kinetic plasma phenomena.
\section{Conclusions}
KEMPIC-3D, a fully electromagnetic three-dimensional particle-in-cell code has been developed for the self-consistent simulation of kinetic plasma phenomena involving the coupled evolution of charged particles and electromagnetic fields. The implementation combines a staggered Yee finite-difference time-domain solver with a relativistic Boris particle pusher, trilinear field interpolation, and charge-conserving current deposition within a modular computational framework suitable for multidimensional plasma simulations.

The numerical implementation has been systematically verified through a hierarchy of benchmark problems addressing both the individual numerical components and the self-consistent particle--field coupling. Independent verification of the electromagnetic field solver confirmed a convergence rate consistent with the expected second-order accuracy of the coupled space-time refinement employed, while the particle pusher accurately reproduced the analytical solutions for relativistic charged-particle motion and exhibited the expected second-order temporal convergence. The self-consistent PIC implementation was subsequently assessed through plasma oscillations, global energy conservation, and discrete charge continuity, demonstrating the numerical accuracy and self-consistency of the particle--field coupling.

The capabilities of KEMPIC-3D were further demonstrated through a representative simulation of nonlinear laser-driven plasma wakefield generation in an underdense plasma. The simulations reproduced the characteristic features of the nonlinear wakefield regime, including the formation of an electron density cavity, the excitation of strong accelerating electric fields, and the expected reduction of statistical noise with increasing numbers of super-particles, while preserving the principal wakefield features over the range of particle discretization considered and without the use of spatial filtering or smoothing.

KEMPIC-3D provides a transparent and extensible computational framework for the investigation of kinetic plasma phenomena. Its modular architecture and direct accessibility of the principal routines facilitate inspection and modification, and further development of the underlying algorithms. This structure enables the incorporation of additional physical and numerical capabilities, including ionization processes, particle collisions, advanced boundary conditions, adaptive particle management, higher-order particle shape functions, and diagnostics tools. Future developments may also include further optimization of existing parallelization strategy to improve computational efficiency for larger multidimensional simulations. These extensions will broaden the applicability of KEMPIC-3D to plasma-based acceleration, laboratory plasma experiments, fusion-related applications, and other problems involving self-consistent kinetic plasma dynamics.

To facilitate reproducibility, independent use, and further development, the KEMPIC-3D source code will be made publicly available through a GitHub repository.
\section{Code availability}
The source code of KEMPIC-3D will be made publicly available through a GitHub repository to facilitate reproducibility, independent use, and further development. Users will be able to download, run, inspect, and modify the source code for research and educational purposes. Researches using KEMPIC-3D in published work are requested to cite the present article. The repository URL will be provided upon publication.


\begin{thebibliography}{00}
\bibitem{Dawson1983}
J.~M. Dawson,
Particle simulation of plasmas,
\textit{Rev. Mod. Phys.} 55 (1983) 403--447.
doi:{10.1103/RevModPhys.55.403}

\bibitem{BirdsallLangdon2004}
C.~K. Birdsall, A.~B. Langdon,
\textit{Plasma Physics via Computer Simulation},
Taylor \& Francis, New York, 2004.

\bibitem{HockneyEastwood1988}
R.~W. Hockney, J.~W. Eastwood,
\textit{Computer Simulation Using Particles},
IOP Publishing, Bristol, 1988.

\bibitem{Fonseca2002}
R.~A. Fonseca, L.~O. Silva, F.~S. Tsung, V.~K. Decyk, W.~Lu,
C.~Ren, W.~B. Mori, S.~Deng, S.~Lee, T.~Katsouleas,
J.-C. Adam,
OSIRIS: A three-dimensional, fully relativistic particle-in-cell
code for modeling plasma-based accelerators,
in: P.~M.~A. Sloot, A.~G. Hoekstra, C.~J.~K. Tan,
J.~J. Dongarra (Eds.),
\textit{Computational Science---ICCS 2002},
Lecture Notes in Computer Science, Vol.~2331,
Springer, Berlin, 2002, pp.~342--351.
doi:{10.1007/3-540-47789-6\(_36\)}

\bibitem{Arber2015}
T.~D. Arber, K.~Bennett, C.~S. Brady, A.~Lawrence-Douglas,
M.~G. Ramsay, N.~J. Sircombe, P.~Gillies, R.~G. Evans,
H.~Schmitz, A.~R. Bell, C.~P. Ridgers,
Contemporary particle-in-cell approach to laser--plasma modelling,
\textit{Plasma Phys. Control. Fusion} 57 (2015) 113001.
doi:{10.1088/0741-3335/57/11/113001}

\bibitem{Derouillat2018}
J.~Derouillat, A.~Beck, F.~Pérez, T.~Vinci, M.~Chiaramello,
A.~Grassi, M.~Flé, G.~Bouchard, I.~Plotnikov, N.~Aunai,
J.~Dargent, C.~Riconda, M.~Grech,
Smilei: A collaborative, open-source, multi-purpose
particle-in-cell code for plasma simulation,
\textit{Comput. Phys. Commun.} 222 (2018) 351--373.
doi:{10.1016/j.cpc.2017.09.024}

\bibitem{Vay2018}
J.-L. Vay, A.~Almgren, J.~Bell, L.~Ge, D.~P. Grote, M.~Hogan,
O.~Kononenko, R.~Lehe, A.~Myers, C.~Ng, J.~Park, R.~Ryne,
O.~Shapoval, M.~Thévenet, W.~Zhang,
Warp-X: A new exascale computing platform for beam--plasma
simulations,
\textit{Nucl. Instrum. Methods Phys. Res. A} 909 (2018) 476--479.
doi:{10.1016/j.nima.2018.01.035}

\bibitem{Lehe2016}
R.~Lehe, M.~Kirchen, I.~A. Andriyash, B.~B. Godfrey, J.-L. Vay,
A spectral, quasi-cylindrical and dispersion-free
particle-in-cell algorithm,
\textit{Comput. Phys. Commun.} 203 (2016) 66--82.
doi:{10.1016/j.cpc.2016.02.007}

\bibitem{VillasenorBuneman1992}
J.~Villasenor, O.~Buneman,
Rigorous charge conservation for local electromagnetic field solvers,
\textit{Comput. Phys. Commun.} 69 (1992) 306--316.
doi:{10.1016/0010-4655(92)90169-Y}

\bibitem{Esirkepov2001}
T.~Zh. Esirkepov,
Exact charge conservation scheme for particle-in-cell simulation
with an arbitrary form-factor,
\textit{Comput. Phys. Commun.} 135 (2001) 144--153.
doi:{10.1016/S0010-4655(00)00228-9}

\bibitem{lopez2025particle}
J.~E. L\'opez and E.~A. Orozco-Ospino,
Particle-in-cell simulations of plasma wakefield formation in microwave waveguides,
\textit{Physics of Plasmas} \textbf{32} (11) (2025).

\bibitem{lopez2026numerical}
J.~E. L\'opez and E.~A. Orozco-Ospino,
Numerical study of electron acceleration by microwave-driven plasma wakefields in rectangular waveguides,
\textit{Physics of Plasmas} \textbf{33} (7) (2026).

\bibitem{Yee1966}
K.~S. Yee,
Numerical solution of initial boundary value problems involving Maxwell's equations in isotropic media,
\textit{IEEE Trans. Antennas Propag.} 14 (1966) 302--307.
doi:{10.1109/TAP.1966.1138693}

\bibitem{Taflove2005}
A.~Taflove, S.~C. Hagness,
\textit{Computational Electrodynamics: The Finite-Difference Time-Domain Method}, 3rd ed., Artech House, Boston, 2005.

\bibitem{Verboncoeur2005}
J.~P. Verboncoeur,
Particle simulation of plasmas: review and advances,
\textit{Plasma Phys. Control. Fusion} 47 (2005) A231--A260.
doi:10.1088/0741-3335/47/5A/017

\bibitem{Boris1970}
J.~P. Boris,
Relativistic plasma simulation—optimization of a hybrid code,
in: \textit{Proceedings of the Fourth Conference on Numerical Simulation of Plasmas},
Naval Research Laboratory, Washington, DC, USA, 1970, pp. 3--67.

\bibitem{EZ}
K.~Steiniger, R.~Widera, S.~Bastrakov, M.~Bussmann,
S.~Chandrasekaran, B.~Hernandez, K.~Holsapple, A.~Huebl,
G.~Juckeland, J.~Kelling, M.~Leinhauser, R.~Pausch,
D.~Rogers, U.~Schramm, J.~Young, A.~Debus,
EZ: An efficient, charge conserving current deposition algorithm for
electromagnetic particle-in-cell simulations,
\textit{Comput. Phys. Commun.} 291 (2023) 108849.
doi:10.1016/j.cpc.2023.108849

\bibitem{Umeda2003}
T.~Umeda, Y.~Omura, T.~Tominaga, H.~Matsumoto,
A new charge conservation method in electromagnetic particle-in-cell simulations,
\textit{Comput. Phys. Commun.} 156 (2003) 73--85.
doi:10.1016/S0010-4655(03)00437-5

\bibitem{Roache1998}
P.~J. Roache,
\textit{Verification and Validation in Computational Science and Engineering},
Hermosa Publishers, Albuquerque, NM, 1998.

\bibitem{Oberkampf2010}
W.~L. Oberkampf, C.~J. Roy,
\textit{Verification and Validation in Scientific Computing},
Cambridge University Press,
Cambridge, 2010.

\end{thebibliography}
\end{document}